\documentclass[aip,graphix,amsmath,amssymb,superscriptaddress,reprint]{revtex4-1}
\usepackage[latin1]{inputenc}
\usepackage{hyperref}
\usepackage{graphicx}
\bibstyle{abbrv-unsrt}
\bibstyle{apsrev}
\usepackage{amssymb,bbm}

\newcommand{\REV}[2]{{#2}}

\newcommand{\dt}{\Delta t}
\newcommand{\dx}{\Delta x}
\renewcommand{\r}{\vec{r}}
\newcommand{\x}{\vec{x}}
\newcommand{\cs}{{c_{\rm s}}}
\newcommand{\fk}{f_{k}}
\newcommand{\fkeq}{\fk^{\rm eq}}
\renewcommand{\c}{\vec{c}}
\newcommand{\ck}{\c_{k}}
\renewcommand{\u}{\vec{u}}
\newcommand{\ueq}{\u^{\,\rm eq}}
\newcommand{\rhor}{{\rho}_{\circ}}
\newcommand{\g}{\vec{g}}
\newcommand{\D}{\mathcal{D}}
\newcommand{\Rj}{\vec{R}_{j}}

\newcommand{\m}{\rm m}
\newcommand{\s}{\rm s}

\newcommand{\pwr}[1]{\times 10\sp{#1}}
\newcommand{\pwrr}[1]{10\sp{#1}}

\renewcommand{\Re}{\ensuremath{\text{Re}}} 

\begin{document}

\title{Quantification of the performance of chaotic micromixers
   on the basis of finite time Lyapunov exponents} 

\begin{abstract}
Chaotic micromixers such as the staggered herringbone mixer developed by
Stroock et al.~allow efficient mixing of fluids even at low Reynolds number by
repeated stretching and folding of the fluid interfaces. The ability of the
fluid to mix well depends on the rate at which ``chaotic advection'' occurs in
the mixer. An optimization of mixer geometries is a non trivial task which is
often performed by time consuming and expensive trial and error experiments. \REV{1-3}{In
this paper an algorithm is presented that applies the concept of
finite-time Lyapunov exponents to obtain a quantitative measure of the chaotic
advection of the flow and hence the performance of micromixers.} By performing
lattice Boltzmann simulations of the flow inside a mixer geometry, introducing
massless and non-interacting tracer particles and following their trajectories
the finite time Lyapunov exponents can be calculated. \REV{1-3}{The applicability of the
method is demonstrated by a comparison of the improved geometrical structure
of the staggered herringbone mixer with available literature data.}
\end{abstract}

\author{Aniruddha Sarkar}
\affiliation{Institute for Computational Physics, University of Stuttgart, 
Pfaffenwaldring 27, 70569 Stuttgart, Germany. }
\affiliation{Department of Applied Physics, Eindhoven University of
Technology, Den Dolech 2, 5600 MB Eindhoven, The Netherlands.}

\author{Ariel Narv\'{a}ez}
\affiliation{Department of Applied Physics, Eindhoven University of
Technology, Den Dolech 2, 5600 MB Eindhoven, The Netherlands.}

\author{Jens Harting}
\affiliation{Department of Applied Physics, Eindhoven University of
Technology, Den Dolech 2, 5600 MB Eindhoven, The Netherlands.}
\affiliation{Institute for Computational Physics, University of Stuttgart, 
Pfaffenwaldring 27, 70569 Stuttgart, Germany. }

\date{\today}

\maketitle

\section{Introduction}
Microfluidic devices have found applications in various scientific and
industrial processes. A typical example is their integration as
important components of chemical and biological sensors.~\cite{30} A
micromixer is a microfluidic device used for effective mixing of different
fluid constituents. It can be used to efficiently mix for example a
variety of bio-reactants such as bacteria cells, large DNA molecules,
enzymes and proteins in portable integrated microsystems with minimum
energy consumption. It is also used in mixing of solutions in chemical
reactions,~\cite{31} sequencing of nucleic acids or drug solution
dilution. Hence, the design of practical and efficient micromixers is a
major research topic in microfluidics,~\cite{12,38} especially in the
development of micro total analysis systems. Over the years various
methods of efficient mixing have been developed and many of those have
been successfully applied in industry.~\cite{26}

The small length scales of the micromixers has a negative impact on mixing as
it results in laminar flows inside the channels. In this flow regime, mixing
is influenced mainly by the process of molecular inter-diffusion.~\cite{2}
Experiments with channels with complex surface topology including grooved
walls have revealed that microscale mixing is enhanced by ``chaotic
advection'', a process which was first reviewed by Aref in 1984.~\cite{3} He
describes how mixing is still possible even at low Reynolds number by repeated
stretching and folding of fluid elements.~\cite{jana-chakravarthy} If
properly applied, this mechanism causes the interfacial area between the
fluids to increase exponentially, which can then lead to an enhanced
intermaterial transport, hence mixing. A comprehensive mathematical
description of the exponential growth of interfacial surfaces can be found in
the book by Ottino.~\cite{5} Mixers were designed in the following years,
which utilize the principle of ``chaotic advection''.~\cite{33} However, it is
important to note that the term ``chaotic'' is used strictly in a Lagrangian
sense.~\cite{PhysFluids.8.5}
 
Depending on the working principle, micromixers can be categorized into
two important types: if energy from an external source is used to drive
the mixing process, then it is termed as ``active mixer''. These external
energy sources can be acoustic bubble induced vibrations, periodic
variation of the flow rate, piezoelectric vibrating membranes, valves, etc.
The external sources are often moving components such as micropumps and
they require advanced fabrication steps.~\cite{Science.25.2007} The second
category of micromixers is based on restructuring the flow profile using
static but sophisticated mixer geometries. These are termed as ``passive
mixer''. Passive micromixers have the advantage of simple fabrication,
easy operation and no elements which can generate heat. The absence of
heating is an important factor for applications to biological studies
where temperature is a sensitive parameter. The mixing length and mixing
time are defined as the distance and time span the fluid constituents have
to flow inside the mixer in order to obtain a homogeneous mixture. An
effective micromixer should reduce the mixing length and mixing time
substantially in order to achieve rapid mixing. A common practice to
design passive micromixers is to create alternating thin fluid lamellae.
These result in an interfacial area that increases with the number of
lamellae rendering the diffusion process more effective and hence allowing
fast mixing.~\cite{22} There are many examples of
bi-lamellation~\cite{23,24} and multi-lamellation~\cite{26} in the
literature on micromixers. However, the drawback of such devices is that
the number of lamellae is generally limited due to the negative impact on
the applied pressure drop caused by the required microstructures inside
the channel. Other examples of passive micromixers include the twisted
pipe mixer,~\cite{9} the superfocus micromixer, where several jets are
made to collide at the focal point of the jets and the three dimensional
serpentine model.~\cite{26}

\begin{figure*}[!t]
{\includegraphics[width=0.8\textwidth]{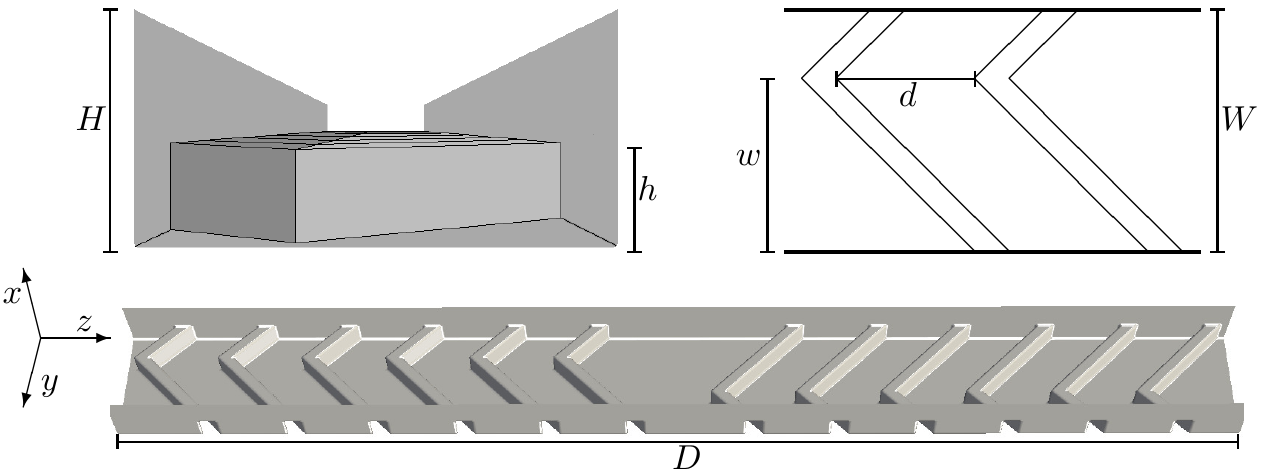}}
\caption{\label{fig_1} A typical example of a SHM geometry as it is used for
  the simulations. The dimensions of this channel are 32$\times$64$\times
  D/\dx$ lattice units, where $D$ depends on the distance between the grooves
  $d$ and the number of grooves per half cycle $n$. $H$ is the height of the
  channel and $w$ is the horizontal length of the long arm. We define the
  height fraction as $\alpha= h/H$, width fraction as $\beta=w/W$, and the
  distance fraction as $\gamma=d/D$. The wall at $x=32$ is not shown in
    the figure, in order to view the inside of the channel.}
 \end{figure*}

The recently developed so-called ``chaotic micromixer'' has gained substantial
interest in the literature since it overcomes some of the drawbacks of
conventional mixers based on muti-lamellation. Such a device consists of
microstructured objects such as ``herringbones'' (see Fig.~\ref{fig_1}). The
staggered herringbone mixer (SHM) was introduced as one of the first
experimental implementations of chaotic micromixers in 2002 by Stroock et
al.~\cite{1} The half cycles of the SHM consist of grooves with two arms which
are asymmetric and unequal in length.  These arms are inclined at an angle of
$45^{\circ}$ \REV{2}{to the wall and $90^{\circ}$ against each other, while}
the pattern interchanges every half cycle of the herringbone. The peculiar
arrangement of the herringbone structure enhances the mixing process by
``chaotic advection'' where the interfacial area between the fluids grows
exponentially in time -- another important advantage over mixers using the
concept of multi-lamellation.  The SHM was used by several authors for studies
related to mixing and analysis of mixing quality. To characterize the mixing
quality of the SHM, Aubin et al.~\cite{40} implemented particle tracking
techniques using the variance of tracer dispersion. Li and Chen report on an
optimization of the SHM using the standard deviation of particle
concentrations. As in the current paper, they used the lattice Boltzmann (LB)
method to model the flow field.~\cite{10} Further, the SHM was used by
Kirtland et al.~\cite{41} to study the mass transfer to reactive boundaries
from three-dimensional micro-channels.

For the development of micromixers it is important to have reliable tools at
hand to quantify their performance. Efficiency and mixing quality have been
studied by various methods in the past. These include the analysis of the
probability density function of the flow profiles, studying the stretching of
the flow field, the Poincar\'{e} section analysis, or the intensity of
segregation as introduced by \REV{2}{Danckwerts in 1952.~\cite{danckwerts-52}}
\REV{3}{In this paper an alternative numerical procedure is
  presented which is tailored for the optimization of chaotic micromixers.} It
is based on LB simulations to describe the flow inside complex mixer
geometries together with a measurement of finite time Lyapunov exponents
(FTLE) as obtained from trajectories of massless tracer particles immersed in
the flow. The LB method can easily handle flows in complex geometries, which
makes this method convenient for flows in advanced microstructures such as the
micromixers the current paper focuses on.  The Lyapunov exponent provides a
quantitative measure of long term average growth rates of small initial flow
perturbations and thus allows a quantification of the efficiency of chaotic
advection.~\cite{45,46} Since the systems of interest are finite and
simulations are limited to a finite time span, the proposed method utilizes
Wolf's method to calculate the FTLE.~\cite{22} 
\REV{4}{While LB simulations of flows in micromixers and FTLE computations to
quantify the degree of chaotic advection have been reported in the literature
before, their combination for 3D performance quantification of realistic
chaotic micromixers has to our knowledge not been published before.}
\REV{3}{The numerical scheme has the potential to assist
  an experimental optimization since geometrical parameters or fluid
  properties can easily be changed without requiring a new experiment.} To
demonstrate its applicability, the scheme is applied to evaluate the
parameters of the staggered herringbone mixer that lead to \REV{1-3}{improved
performance}.

In the following sections the lattice Boltzmann method which is applied to
simulate the fluid flow and the Wolf's algorithm from which the finite
time Lyapunov exponents are obtained are described. Finally, the numerical
results and conclusions are given.

\section{Simulation Method} 
\label{SIMMETH}
The lattice Boltzmann method is used to describe the fluid flow. The LB method
is a simplified approach to solve the Boltzmann equation in discrete space,
time and with a limited set of discrete velocities.~\cite{44} The Boltzmann
equation, given as
\begin{equation} 
\partial_{t} f + \c \cdot \nabla f = \Omega(f),
\end{equation}
represents the evolution of the velocity distribution function by molecular
transport and binary intermolecular collisions.  $f(\r,\c,t)$ represents the
distribution of velocities in continuous position and velocity space, $\r$ and
$\c$ respectively. In the LB approach the position $\x$ at which $f(\x,\ck,t)$
is defined, is restricted to a discrete set of points on a regular lattice
with lattice constant $\dx$, implying that space is discretized. The velocity
is restricted to a set of velocities $\ck$ implying that velocity is
discretized along specific directions. $\dt$ denotes the discrete time step.
The model we adopt is a D3Q19 model which is a 3 dimensional model with 19
different velocity directions, $k = 0,1,...,18$.~\cite{EurophysLett.17.479}
The right hand side of the above equation represents the collision operator
which is simplified to the linear Bhatnagar-Gross-Krook (BGK) form.~\cite{27}
In a discretized form the BGK operator is written as
\begin{equation}
\Omega_k = \omega( \fkeq (\x,t) - \fk (\x,t)).
\end{equation} 
Here, $\omega$ is the reciprocal of the relaxation time of the system
controlling the relaxation towards the Maxwell-Boltzmann equilibrium
distribution $\fkeq (\x,t)$. By considering small velocities and constant
temperature, a discretized third order Taylor expansion of the above
equilibrium distribution function can be written as
\begin{equation}
\begin{split}
  \fkeq (\x,t) = & \zeta_k \frac{\rho}{\rhor} 
  \Bigg( 1 +
  \frac{ \ck \cdot \ueq }{ \cs^2} + \frac{ ( \ck \cdot \ueq )^2}{ 2 \cs^4 } -
  \frac{ \ueq \cdot \ueq }{2 \cs^2 } \\
  &  + \frac{( \ck \cdot \ueq )^3}{ 6 \cs^6 } 
  - \frac{ ( \ck \cdot \ueq )( \ueq \cdot \ueq ) }{ 2 \cs^4 } \Bigg),
\end{split}
\end{equation} 
where $\fkeq$ denotes the equilibrium distribution function corresponding to
the velocity vector $\ck$, $\zeta_k$ are the lattice weights, $\rho$ is the
density, $\rhor$ a reference density, and $\cs = (1/\sqrt{3}) \dx / \dt $ is
the speed of sound. $\ueq$ is the equilibrium velocity of the fluid, which is
shifted from the mean velocity by an amount{ $\g / \omega$ }under the
influence of a constant acceleration $\g$. The evolution of the LB process
takes place in two steps: the collision step where the velocities are
redistributed along the directions of the lattice and the propagation step by
which they are displaced along these directions. These discrete LB steps are
implemented by the equation
\begin{equation} 
\label{eqn_3}
\fk (\x + \dt \ck, t + \dt) - \fk (\x,t) =
-\omega \dt ( \fkeq (\x,t) -  \fk (\x,t) ),
\end{equation} 
which gives the dynamic evolution of the distribution function and is referred
to as a discretized Boltzmann kinetic equation. The macroscopic fluid density
is given by
\begin{equation} 
\rho (\x,t) = \rhor \sum_{k} \fk (\x,t) 
\end{equation} 
and the macroscopic fluid velocity in the presence of external forcing is
given by~\cite{JSTATMECH.2010.P11026}
\begin{equation} 
\u (\x,t) =
\frac{ \rhor }{ \rho(\x,t) } 
\sum_{k} \fk (\x,t) \ck  - \frac{\dt}{2}\g.
\end{equation}
It can be shown by a Chapman-Enskog expansion that the macroscopic fields $\u$
and $\rho$ from the above equations fulfill the Navier Stokes equation in the
low Mach number limit and for isothermal systems.~\cite{44} In order to
simulate a fluid flow through microchannels, periodic boundary conditions are
implemented along the $z$ direction (see Fig.~\ref{fig_1}) and no-slip bounce
back boundary conditions are imposed at the channel walls.  The experimental
setup of Stroock et al.  consists of a number of repeated cycles ($\sim$15)
consisting of two asymmetric half cycles each. By simulating one full cycle
only and applying periodic boundary conditions we can reduce the required
computational effort substantially. \REV{3}{If the system is in steady state,
  we do not expect any substantial differences in the flow field in different
  cycles. Therefore, this approach is valid. If, however, the experimental
  system would not be periodic, such a simplification would not be allowed and
  the full geometry would have to be simulated.}

\begin{figure*}[t!]{
\begin{tabular}{ll}
a)  & b)\\
  \includegraphics[width=0.45\textwidth]{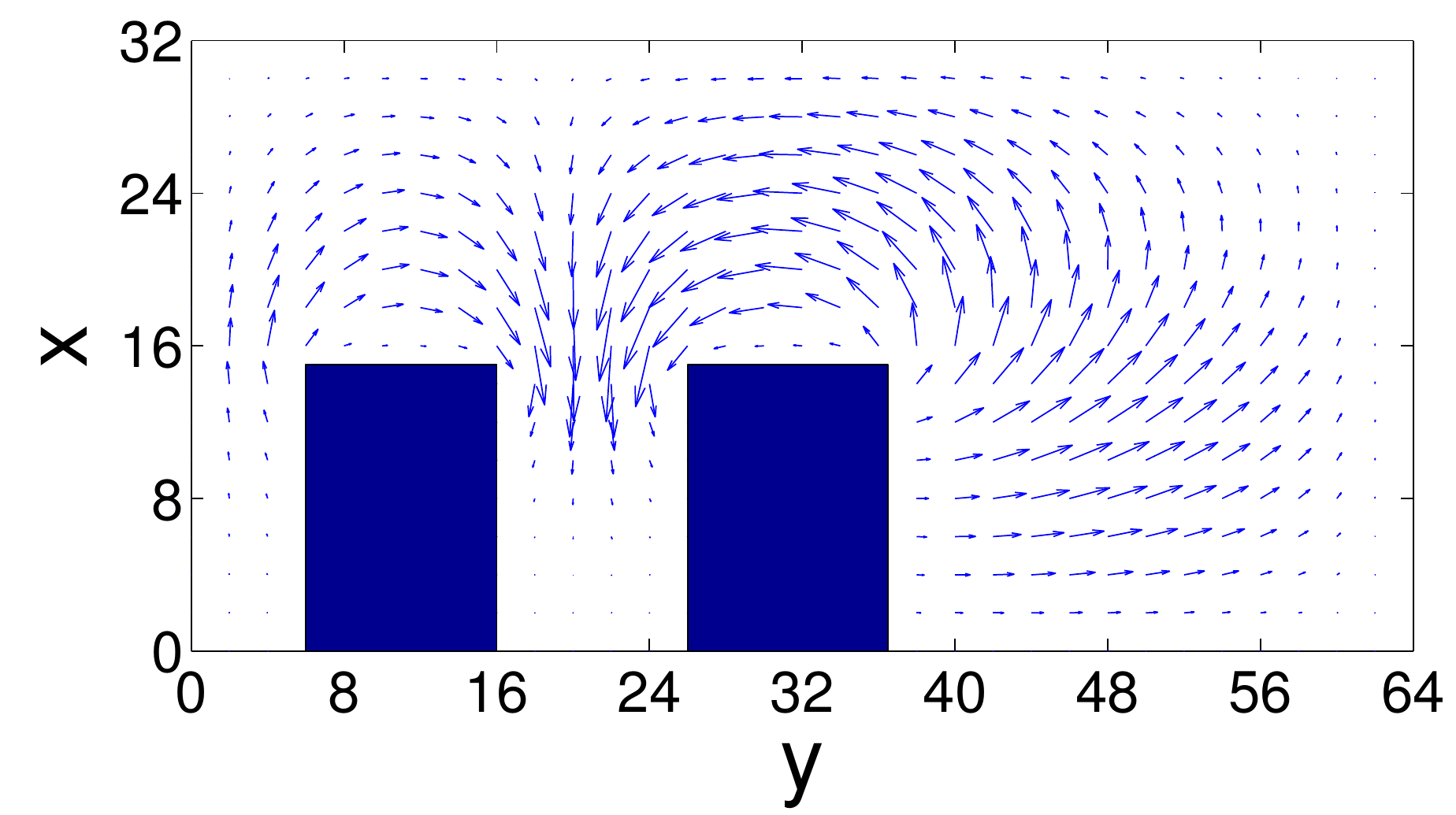}
&
  \includegraphics[width=0.45\textwidth]{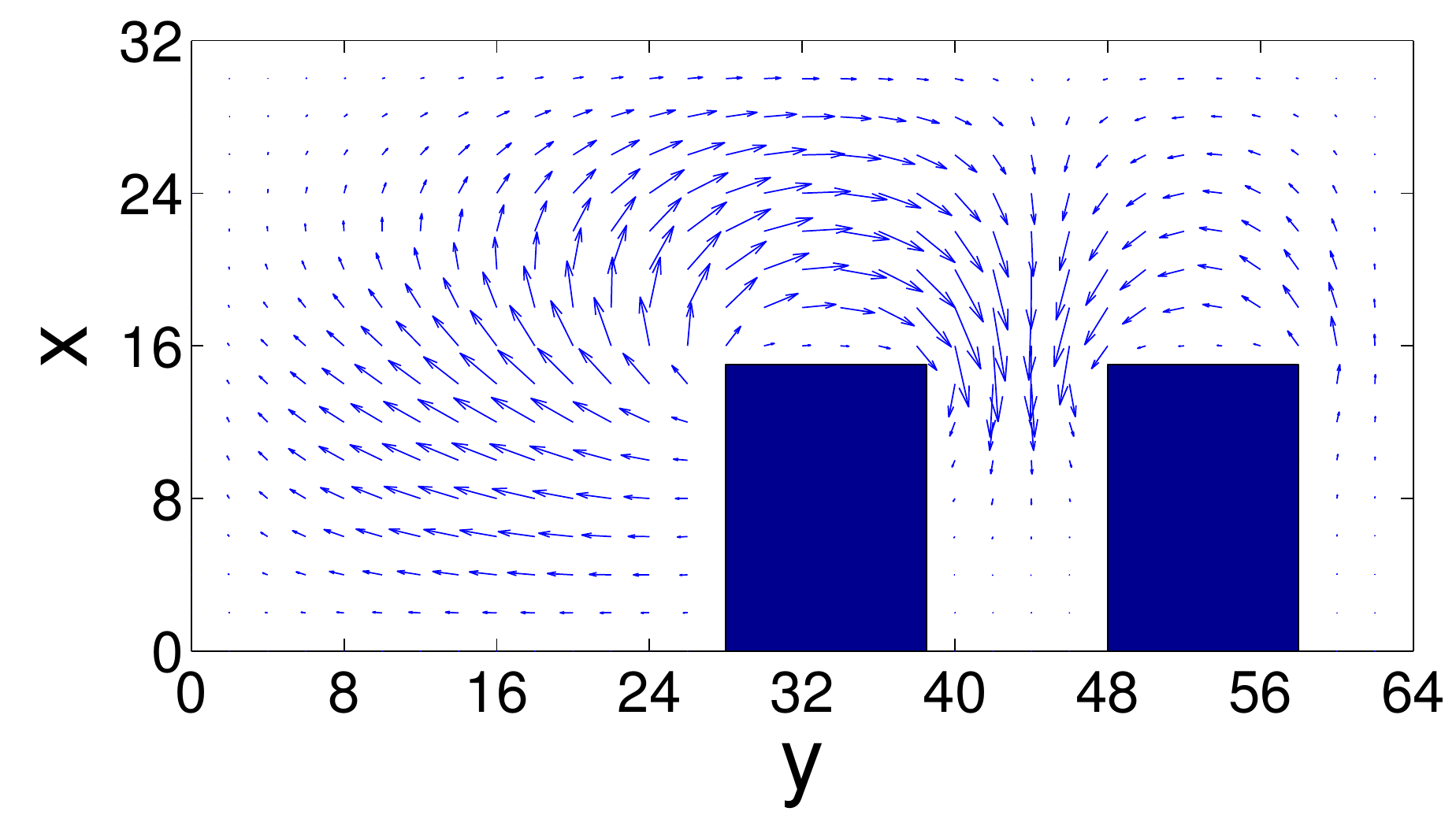} \\
\multicolumn{2}{l}{c)} \\
\multicolumn{2}{c}{
\includegraphics[width=0.9\textwidth]{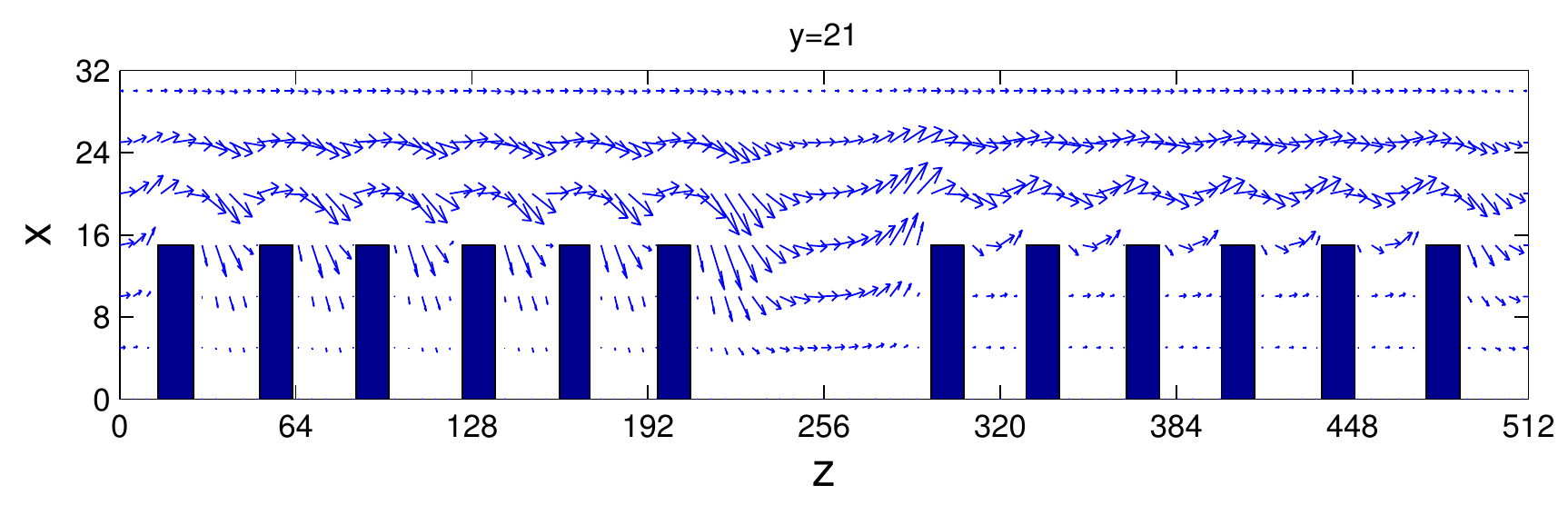}
} 
\end{tabular}
\caption{\label{profile} The velocity field perpendicular to the flow
  direction at $z=30$ and $z=460$ is shown in insets a) and b).  The figures
  depict the circulating motion of the flow which is triggered by the
  asymmetry of the herringbone shaped surface structures. Inset c) shows the
  velocity field in flow direction at $y=21$. Due to the choice of $\beta =
  0.66$ this value corresponds to a position close to the tips of the
  herringbones in case of the first half cycle and to a case in the center of
  the long arm in the case of the second half cycle. As can be observed the
  velocity vectors point strongly downwards within the first half of the
  channel, but mostly upwards in the second half.  } }
\end{figure*}

We simulate a fluid which is hydrodynamically similar to water, flowing inside
a SHM with a cross section of $96\,\mu\m \times 192\,\mu\m$. The length of the
channel is $1536\mathrm{\mu m}$, but can be varied in order to always
accommodate a full cycle of the herringbone structure. For computational
efficiency we have chosen a lattice resolution of $\dx = 3\,\mu\m$ resulting
in a fixed cross section of $32 \dx \times 64 \dx$ and system lengths of the
order of $512 \dx$.  Such a relatively low resolution is sufficient to
properly resolve the flow field as can be observed from Fig.~\ref{profile},
where the velocity field perpendicular to the flow is shown at positions
within different half cycles (a,b). Fig.~\ref{profile}c) shows the profile in
flow direction.  While the first two subfigures nicely demonstrate the
swirling motion of the flow, the third one depicts how the flow penetrates
between the grooves and stays mostly unaffected close to the upper boundary of
the channel.  Previously, we have shown that a well resolved velocity field
with less than 4 percent error as compared to the analytical solution can be
obtained even for a resolution of 6-8 lattice nodes in the case of a 3D
rectangular Poiseuille flow.~\cite{JSTATMECH.2010.P11026} We further
demonstrated that flow over random rough surfaces can be well resolved even if
the smallest obstacles are only described by 2-4~lattice
units.~\cite{Kunert.2007,Kunert.2010} In the LB method, the kinematic
viscosity is related to the discrete time step through the expression $\nu =
\cs^2 ( 1 / \omega - \dt / 2 )$.  Since ${ \omega \dt }$ is chosen to be $1$
to minimize artefacts due to the mid-grid bounce back boundary conditions and
the simulated fluid has the kinematic viscosity of water, $\nu =
\pwrr{-6}\,\m^{2}\,\s^{-1}$, this implies for the current choice of $\dx$ that
$\dt = 1.5 \pwr{-6}\,\s$ and $\cs = 1.15\,\m\,\s^{-1}$. When the magnitude of
$\g$ is $1.2\pwr{-3}\,\m\,\s^{-2}$ along the $z$ direction, the average
steady state velocity of the system is $u\approx 6.0\pwr{-3}\,\m\,\s^{-1}$
which corresponds to a subsonic flow. The Reynolds number $\Re = u \, L / \nu$
of the flow is $\approx 1.3$, where $L = \sqrt{ H^2 + W^2 }$ is the
characteristic length of the channel. One set of simulations is obtained for
$\g$ being $0.4\pwr{-3}\,\m\,\s^{-2}$ which corresponds to $\Re \approx 0.4$.

When the flow simulation has reached its steady state, $P=1000$ massless and
non-interacting tracer particles are introduced at fluid nodes in the $z=0$
plane and then their velocities are integrated at each time step. To calculate
the FTLE from particle trajectories a group of five particles forms four
pairs, with every 5th particle placed at the center and the remaining ones
being placed at the four nearest off-diagonal neighboring LB lattice
sites. The particle at the center traces a fiducial orbit. With this
arrangement we are able to follow 800~particle pairs by using only
1000~particles.

The trajectories are obtained by integrating the vector equation of motion   
\begin{equation}
 \frac{ {\rm d} \Rj }{ {\rm d}t} = \u (\Rj), \quad j=1,...,P
\end{equation}
where $\Rj $ denotes the position vector of an individual tracer particle.
The velocity $\u (\Rj)$ is obtained from the discrete LB velocity
field through a trilinear interpolation scheme. 

``Chaotic'' systems in general have the important feature that two nearby
trajectories diverge exponentially in time. The rate at which these
trajectories diverge can be related to the ability of the flow field to create
conditions for chaotic mixing. The Lyapunov exponent is a possible measure of
the performance of a micromixer as it is related to the rate of stretching of
the fluid elements. It is defined by
\begin{equation} 
 \lambda_{\infty} = \lim_{t\to\infty} \frac{1}{t} \ln\frac{ \D (t) }{ \D (0) },
\end{equation}
where $\D (t)$ is the distance between two trajectories at time $t$ which
evolve from an initial separation $\D (0)$. $\lambda_{\infty}$ gives the value
of the Lyapunov exponent as $t$ tends to infinity. Due to the finite size of
any microfluidic system it is not possible to implement this definition in a
simulation code to study the performance of a micromixer Also, when two
trajectories separate from each other, this definition does not allow to
understand the ongoing stretching and folding dynamics of the flow. A
quantitative measure of the mixer performance based on the Lyapunov exponent
can be obtained by using the FTLE instead of the previous
expression.~\cite{25,32} The FTLE takes the dynamical process more completely
into account and provides a numerically implementable scheme for
quantification of the performance of mixers. The FTLE is defined as~\cite{34}
\begin{equation}
\lambda_{\rm FTLE} = \frac{1}{\delta t}   \ln\frac{\D (t+\delta t)}{ \D (t) } ,
\end{equation} 
where $t$ is any particular instant of time and $\delta t$ is a finite time
after which the FTLE is measured. The same process is repeated over $N$ times,
where $N$ is a large number denoting the number of times the FTLE is evaluated
from trajectories of particle pairs.  \REV{2}{Amon et al.} named the FTLE as
finite time Lagrangian Lyapunov exponent.~\cite{PhysFluids.8.5} The
convergence of the average FTLE to the Lyapunov exponent for large $N$ is
discussed in the paper by Tang and Boozer,~\cite{32}
\begin{equation}
 \lim_{N \to \infty} \langle\lambda_{\rm FTLE}\rangle_N = \lambda_{\infty}.
\end{equation} 

\begin{figure}[h]
\centerline{\includegraphics[width=1.0\columnwidth]{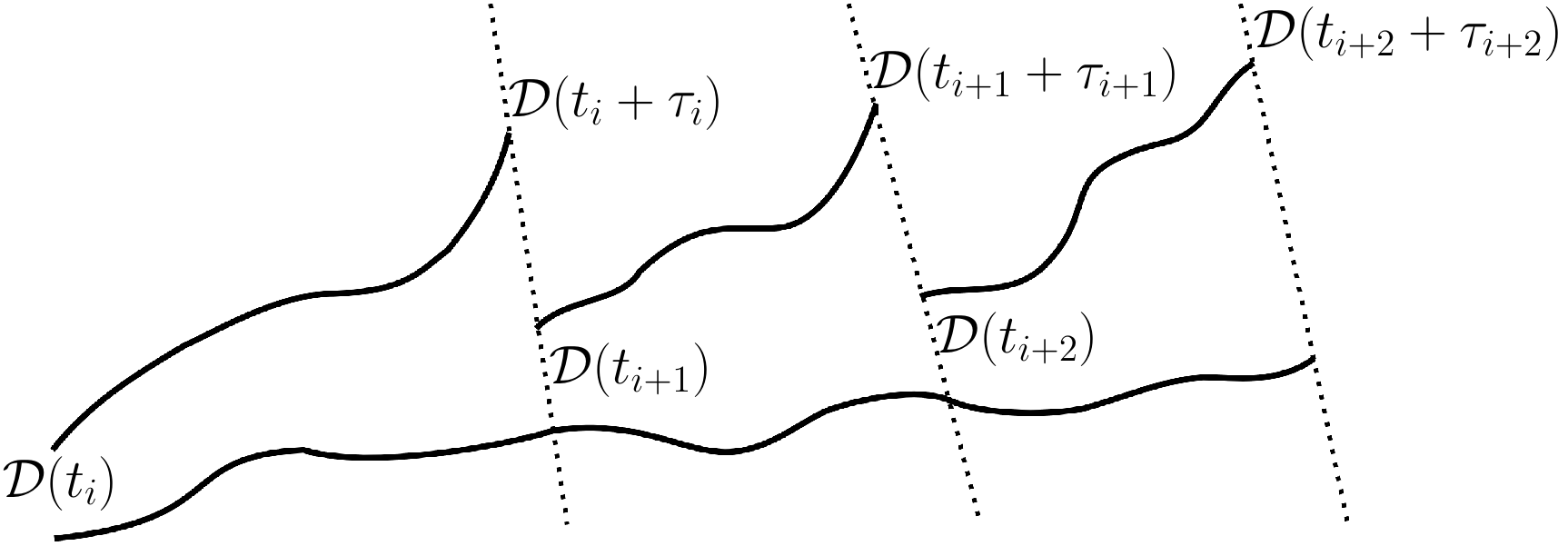}}
\caption{\label{fig_2}A schematic representation of Wolf's method.  $\D (t_i)$
  is the distance between a particle pair at an arbitrary time $t_i$.  If the
  distance is greater than a maximum distance then one of the particles is
  replaced near to the other particle along the line of separation.}
\end{figure}

Wolf et al. suggested a method to calculate the FTLE from a set of
experimental data which is well applicable to our simulations.~\cite{22,4}
\REV{3}{The key idea of the method is to monitor the distance between
  particles forming a pair and to renormalize it by moving back one of the two
  particles towards the other one in case they have separated more than a
  given threshold distance. The FTLE is then computed from the sum of
  individual Lyapunov exponents measured between replacements.}  The method
has been verified on systems with known Lyapunov spectra and exact results
have been achieved. Following Wolf's approach, we implement the following
equation to quantify the mixer performance on the basis of the average FTLE as
\begin{equation}
\langle\lambda\rangle_{N} = \frac{1}{N} \sum_{i=0}^{N-1}
\frac{1}{\tau_{i}}\ln\frac{ \D (t_{i}+ \tau_{i})}{ \D ( t_{i} ) },
\end{equation}
where $t_{i}$ is the $i$th time when a FTLE is evaluated, $ \D (t_{i} +
\tau_{i})$ and $\D (t_{i})$ are the distance between particle pairs at time
step $t_{i}+\tau_{i}$ and $t_{i}$, respectively. $\tau_{i}$ is the number of
time steps which a pair of particles take until the next replacement and its
magnitude could be different for each measure. $N$ is the total number of
replacements made until time $t$ when $\langle\lambda\rangle_{N}$ is
evaluated. If $\langle\lambda\rangle_{N}$ has a positive and non-zero value
the particles separate from each other at an exponential rate. These
particle pairs are initially very close to each other and evolve in time. \REV{3}{If
the separation between the pair is greater than a maximum distance, the distance between the particles is
re-adjusted to the initial distance $\D (t_{0})$. 
For the implementation of the scheme, for every particle pair one of
the trajectories is chosen as the fiducial path, while the position of the
other particle is replaced if the distance becomes larger than the threshold
value. This is schematically represented in Fig.~\ref{fig_2}. 
The choice of
the maximum distance is based on a variation of it together with a
comparison of the obtained FTLE: if it is chosen too large, many tracer
particles hit the channel walls and do not separate any further resulting in a
too low value of the FTLE. If it is chosen too small, the tracers do not have
a sufficient amount of time to separate sufficiently so that an exponential
increase of the distance cannot be detected. The chosen value $H/2$ has been
found to be a good compromise between the two extreme cases.
In order to avoid errors due
to orientation one of the particles of a pair is placed along the line of
separation. In case a replacement point cannot be found due to a wall node
present at the location, a nearby fluid node is selected as the replacement
point. If even such points cannot be found since all surrounding nodes are
surface nodes, the replacement is postponed until a suitable replacement is
possible.} 

The following section presents how FTLE can be utilized for an optimization
strategy for chaotic micromixers. As an example, the influence of different
parameters which directly affect the performance of the SHM is evaluated. These
are the ratio of the height of the grooves to the height of the channel
$\alpha$, the ratio of the horizontal length of the long arm to the channel
width $\beta$, the ratio of distance between the grooves to the length of the
channel $\gamma$ and the number of grooves per half cycle $n$. The width of the
grooves is kept fixed at $24\,\mu\m$ for all simulations. Fig.~\ref{fig_1}
provides a pictorial representation of these parameters. 

\section{Results and discussion}
The performance of the SHM is studied by varying four geometrical parameters. While keeping all other
parameters fixed \REV{1,3}{($\gamma=$0.089, $\alpha=$0.2, $n$=10)}, the width fraction \REV{2}{($\beta=w/W$)} is varied within the range of
\REV{1}{0.5} and 0.82 and the distance fraction \REV{2}{($\gamma=d/D$)} from 0.04 to 0.11. Then,
the number of grooves per half cycle ($n$) is varied from 2 to 10 and the
height fraction \REV{2}{($\alpha=h/H$)} from 0.125 to 0.343. 
\REV{1-3}{The
optimization of the SHM as presented here is meant to demonstrate the
feasibility of the algorithm only since several optimization studies of the SHM
are already available in the literature. Therefore our optimization is reduced
to a limited variation of the four dimensional parameter space. This surely
leads to a local optimum of the geometrical parameters, but it is not assured
that the global optimum has been found. However, as shown below, our parameters
corespond to the optimal ones found by other groups suggesting that our local
optimum coincides with the global optimum.}

\begin{figure}[h]
 \centerline{\includegraphics[width=1.0\columnwidth]{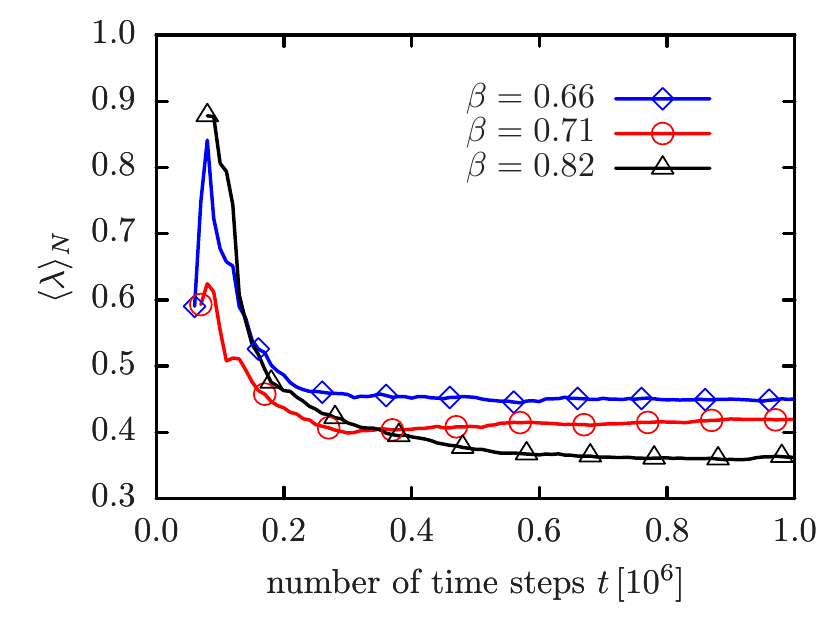}}
 \caption{\label{fig_3} The average FTLE for 1,000 particles is shown
versus the time steps for $\Re = 1.3$. After $6.0\pwr{5}$ time
steps, $\langle\lambda\rangle_N$ is converged and the average of this
converged value is denoted by $\lambda$. The value of $\lambda$ depends on
the width fraction $\beta$ as it is further investigated below.}
\end{figure}

\begin{figure}[t]
 \centerline{\includegraphics[width=1.0\columnwidth]{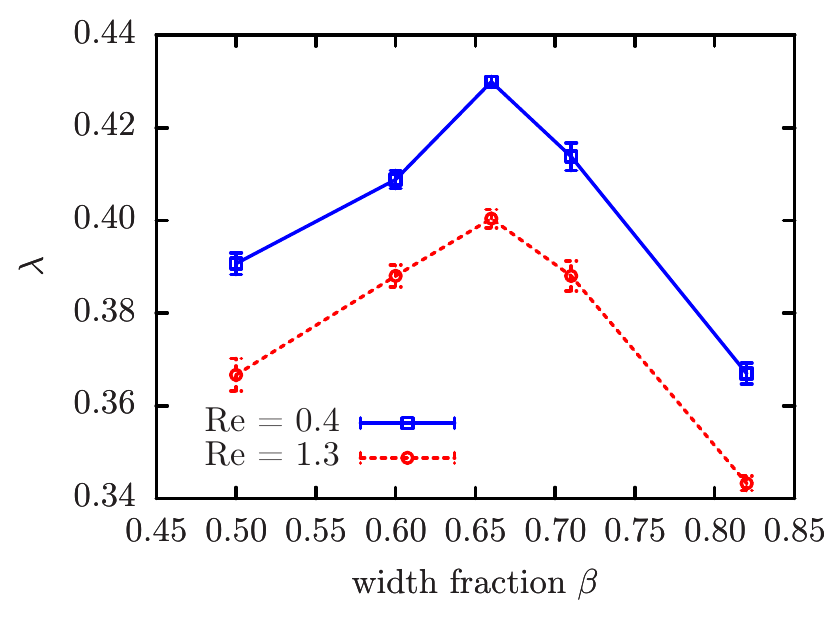}}
 \caption{\label{fig_4} The variation of the maximum averaged finite time
Lyapunov exponent $\lambda$ with different width fraction $\beta$, for two
different Reynolds numbers. The data indicates that the maximum $\lambda$
can be obtained for a width fraction of $\beta = 2/3$. Error bars are given by
the standard deviation of the mean value of $\langle\lambda\rangle_N$.}
\end{figure}

\begin{figure}[h]
 \centerline{\includegraphics[width=1.0\columnwidth]{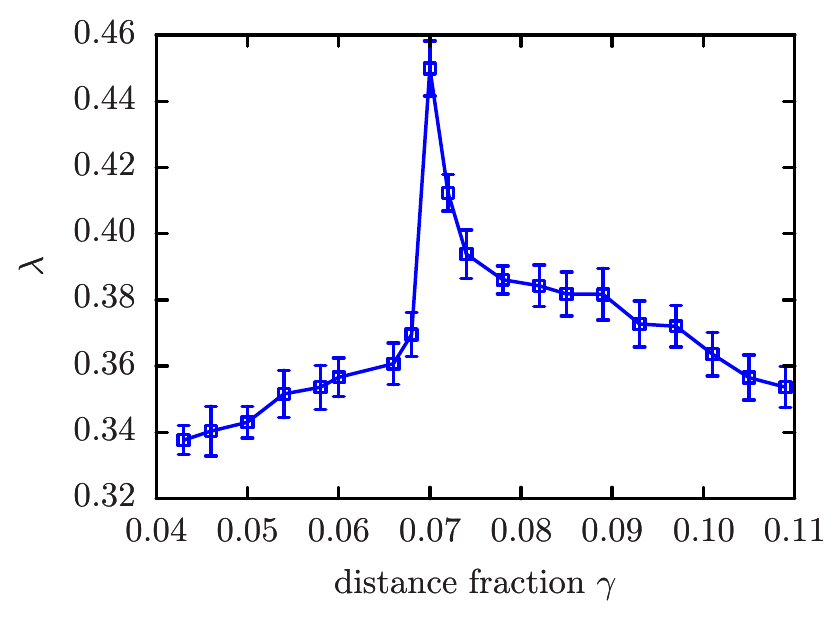}}
 \caption{\label{fig_6} The variation of $\lambda$ versus the distance
   fraction $\gamma$ for $\Re=1.3$ and $\beta=2/3$. The FTLE rises
   with the increase of $\gamma$ until it reaches a distinct peak. After that
   the data shows a decrease, indicating that the optimized performance of the
   micromixer is at a groove distance $\gamma=0.07$. The error bars are
   given by the standard deviation from the mean.}
\end{figure}

\begin{figure}[h]
 \centerline{\includegraphics[width=1.0\columnwidth]{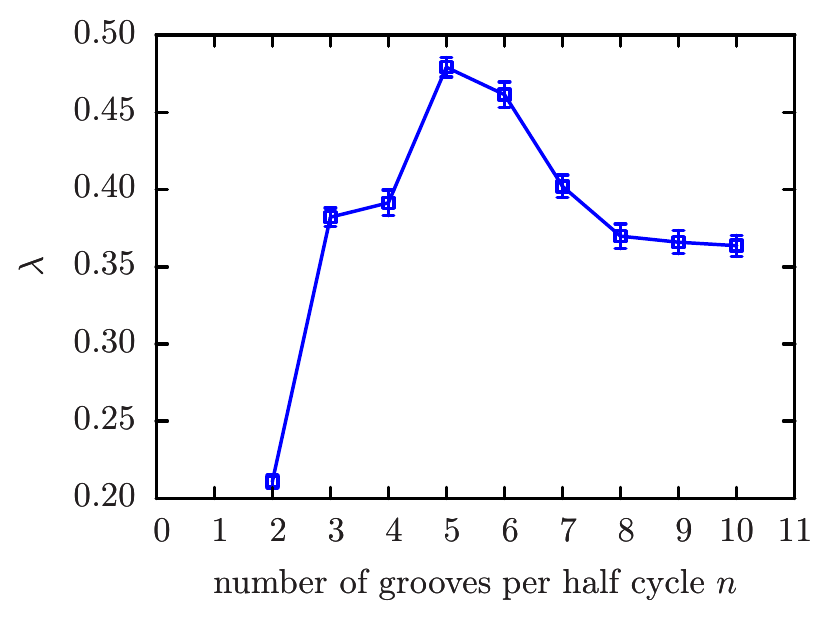}}
 \caption{\label{fig_7} The variation of $\lambda$ with the number of grooves
   per half cycle ($n$) is shown. It can be observed that the SHM with $n=5$
   performs best. Error bars indicate the standard deviation from the mean
   values.}
\end{figure}

\begin{figure}[!h]
 \centerline{\includegraphics[width=1.0\columnwidth]{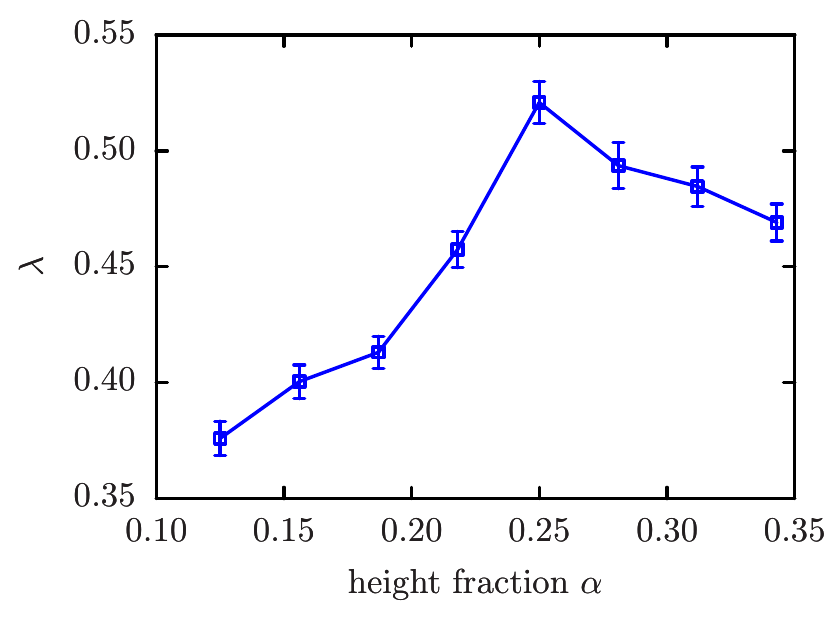}}
 \caption{\label{fig_8} The variation of the average converged FTLE with the
   height fraction ($\alpha$). The data indicates that the maximum FTLE can be
   obtained for $\alpha = 0.25$. Error bars indicate the standard deviation
   from the mean values.}
\end{figure}

Fig.~\ref{fig_3} depicts simulated $\langle\lambda\rangle_N (t)$ for different
width fractions $\beta=0.66$, $0.71$, and $0.82$. The Reynolds number is kept
fixed at $\Re =1.3$ and $\langle\lambda\rangle_N (t)$ is obtained from tracing
the trajectories of 1000~particles. From Fig.~\ref{fig_3} it can be observed
that in each case $\langle\lambda\rangle_N$ fluctuates before finally
converging to a particular value after $\sim 6.0\pwr{5}$ time steps. All
further simulations are run until the FTLE have thoroughly converged. The
effect of the geometry can be measured by comparing the average of the
converged FTLE which is denoted by $\lambda$.  The error bars in
Figs.~\ref{fig_4} to~\ref{fig_8} are given by the standard deviation of the
data from the point where it has converged.

When one of the parameters is varied it is taken care that the other
parameters remain unchanged, because we are interested in the dependence of
the individual parameters on the performance of the SHM.  Fig.~\ref{fig_4}
shows the variance of $\lambda$ and as such the performance of the SHM with
respect to the parameter $\beta$ for two different Reynolds numbers, $\Re =
0.4 $ and 1.3. \REV{3}{Larger values are difficult to obtain due to the limits of the LB method or would include a substantial increase in computing time.} Due to the symmetry of the mixer geometry, only values for
$\beta\ge0.5$ are shown. The datasets peak at $\beta = 2/3$, which implies
that the degree of chaotic advection is maximized for this particular value of
the width fraction $\beta$. According to the units chosen above this
corresponds to $w = 130\,\mu\m$. The curves for the different Reynolds numbers
depict that changing the driving force of the fluid does influence the
absolute value of $\lambda$, but has no influence on the general shape of the
curve.  Similar studies of the \Re\ dependence for other geometrical parameters
and various different driving forces confirm this behavior. Our findings are
consistent with the original experimental work of Stroock et al.~\cite{1} as
well as 2D numerical optimizations by Stroock and McGraw.~\cite{29} The latter
study how the so-called heterogeneity factor $I =\frac{
  \sqrt{\langle(C-\langle C \rangle)^{2}\rangle} } { \langle C \rangle}$ of a
dye concentration $C$ varies with the number of cycles in the mixer. In both
publications it is found that $\beta=2/3$ generates a maximum swirling motion
of the fluid.


For the next set of simulations $\beta$ is fixed at the optimized value of
$2/3$ and the distance fraction $\gamma$ is varied from 0.04 to 0.11.
Since it is observed from the previous results that the optimal value is
independent of the Reynolds number, the following simulations
are performed at a constant $\Re=1.3$. The average value
of converged FTLE for different distance fractions is shown
in Fig.~\ref{fig_6}. It can be observed that after a moderate increase of
$\lambda$ with $\gamma$, the curve has a sharp peak at $\gamma = 0.07$, which
corresponds to a value of $d=105\,\mu\m$ for the current choice of
$\dx$. Afterwards, $\lambda$ decreases in a similar fashion as for
small $\gamma$, but still at higher absolute values. A possible
explanation is as follows: when the grooves are very close to each other,
they create ``dead spaces'', i.e. regions in the micro-channel where the
fluid gets trapped and cannot move freely. With increasing the distance
between the grooves, the transverse component of the velocity increases,
hence ``chaotic advection'' is enhanced resulting a large value for
$\lambda$. On the other hand, if the distance between grooves is too
large, the mixer behaves like a plain channel without any chaotic
advection component. The maximum in Fig.~\ref{fig_6} is then given by the
optimal interplay of these two effects.

After having optimized the values for $\beta$ and $\gamma$, the number of
grooves per half-cycle $n$ is varied from 2 to 10. It can be understood from
Fig.~\ref{fig_7} that a variation of $n$ has the largest impact on the
performance of the mixer as compared to $\beta$ or $\gamma$. For the current
setup, by variation of $n$ it is possible to change the value of $\lambda$ by
a factor of 2.3 as compared to 1.2 for $\beta$ and 1.3 for $\gamma$.
Fig.~\ref{fig_7} clearly shows that a staggered herringbone mixer with $n=5$
performs best. The existence of a maximum performance in dependence on the
number of grooves can be explained by the fact that interplay between
advection in flow direction and the swirling motion needs to be optimized for
a well performing mixer. If the number of grooves is too small, the flow field
is not sufficiently rotated when flowing through a half cycle. The change of
direction of the swirling motion at the beginning of the following half cycle
does not result in a relevant distortion of the trajectories then. If the
number of grooves is too large, however, the fluid might perform one or more
full rotations and come back to its original position before entering the next
half cycle. An optimized value for $n$ therefore depends on the ratio of
optimum rotation to the frequency at which the distortions at the end of a
half cycle occur.  Similar to the work presented in the current paper, Li and
Chen performed LB simulations and used tracers to follow the flow
field.~\cite{10} They, however, quantify mixing by computing the standard
deviation of the local tracer concentration and conclude that SHM with $n=5$
or $n=6$ perform best. Even though this result is in agreement with our
finding, the FTLE analysis clearly shows that the channel with $n=5$ performs
better than the one with $n=6$.

The final parameter to be \REV{1-3}{varied} is the ratio of the half depth of the
grooves to the height of the channel $\alpha$.  Fig.~\ref{fig_8} shows the
average value of the converged Lyapunov exponents for different $\alpha$
between 0.125 and 0.343.  After a strong increase of $\lambda(\alpha)$, the
data shows a maximum at $\alpha = 0.25$. For the units chosen above this
corresponds to a groove depth of $24\,\mu\m$.  For larger $\alpha$ the value of
$\lambda$ decreases again. Our result is similar to the original experimental
analysis of Stroock et al.~\cite{1} Again, an argument can be found for the
existence of an optimal value for the ratio between groove depth and system
height: if the grooves are too shallow, they are not able to generate the
swirling motion required for chaotic advection. On the contrary, if the
grooves are too deep the flow is not able to fully penetrate the
grooves. Further, the volume between grooves and top surface becomes so small
that the swirling motion cannot develop anymore.

In this section we have demonstrated that a numerical scheme based on the LB
method for the flow in complex mixer geometries together with Wolf's method to
calculate FTLE from trajectories of passive tracers are a powerful tool for
the quantification of chaotic mixing. Without loosing generality, a
\REV{1-3}{limited} optimization study was performed for the particular example of the SHM.  The
optimal parameters $\alpha=0.25$, $\beta=2/3$, $\gamma=0.07$ and $n=5$
have been found which is in good agreement with known experimental and
numerical literature data.

\section{Conclusion}
Mixing at the microscale can be efficient if a large interface between fluids
is provided. This can be obtained by passive chaotic micromixers utilizing
repeated stretching and folding of the fluid interfaces.  The performance of
such mixers depends on the rate at which ``chaotic advection'' of the fluid
takes place. For the development of efficient chaotic micromixers it is
mandatory to understand the underlying transport processes as well as their
dependence on the geometric structure of the microfluidic device. In this
paper we have demonstrated an efficient numerical scheme which allows the 
quantification of the performance of a
micromixer. The scheme is based on a LB solver to describe the time
  dependent flow field in complex mixer geometries combined with Wolf's
method to compute FTLE from passive tracer trajectories.  We have demonstrated
the applicability of the quantification method by applying it to \REV{1-3}{find an optimal geometrical configuration of the
SHM, but the scheme should be generally applicable to any chaotic mixer}. By performing a systematic variation of the relevant geometrical
parameters we obtained a set of optimal values which is consistent with
literature data published by others. However, those data was obtained from
experiments or numerical simulations not taking the chaotic nature of the
tracer trajectories into account. The method presented here, however
  allows a quantification and optimization of the mixing performance by
  investigation of the underlying flow profiles.
\REV{2}{Further work could include the optimization of other chaotic
micromixer geometries and the introduction of multiple fluids to the problem.
For the latter, the lattice Boltzmann method offers a number of possibilities
to simulate multicomponent flows and is thus a well suited candidate.} 

\begin{acknowledgments}
The authors thank F. Janoschek, F. Raischel, G.J.F. van Heijst, and M.
Pattanty\'{u}s-\'{A}brah\'{a}m for fruitful discussions. This work was
financed within the DFG priority program ``nano- and microfluidics'', the
DFG collaborative research center 716, and by the NWO/STW VIDI grant of
J. Harting. We thank the J\"{u}lich Supercomputing Center and the
Scientific Supercomputing Center, Karlsruhe for providing the computing
time and technical support for the presented work.
\end{acknowledgments}


\end{document}